\documentclass[prd,aps,floats,twocolumn,nofootinbib,superscriptaddress]{revtex4-1}
\usepackage{slashed}
\usepackage{mathtools}
\usepackage{amsfonts}
\usepackage{amssymb}
\usepackage{epsfig}
\usepackage{rotating}
\usepackage{url}
\usepackage{times}
\usepackage{color}
\usepackage{bm}
\usepackage{xcolor,colortbl}
\usepackage{hyperref}
\usepackage[alsoload=hep]{siunitx}
\usepackage{enumitem}
\usepackage{fancyhdr}
\usepackage{anyfontsize}
\usepackage{wasysym}
\usepackage{empheq}

\newcommand{\nn}{\nonumber}
\newcommand{\ph}{\phantom}
\newcommand{\eps}{\epsilon}
\newcommand{\be}{\begin{equation}}
\newcommand{\ee}{\end{equation}}
\newcommand{\bea}{\begin{eqnarray}}
\newcommand{\eea}{\end{eqnarray}}
\newcommand{\bare}{\bar{e}}

\newcommand{\bd}{\bar{D}}

\begin{document}


\title{Gravity waves in parity-violating Copernican Universes}
\author{Stephon Alexander}
\email{stephon\_alexander@brown.edu}
\affiliation{Department of Physics, Brown University, Providence, RI, 02906, USA}
\author{Leah Jenks}
\email{leah\_jenks@brown.edu}
\affiliation{Department of Physics, Brown University, Providence, RI, 02906, USA}
\author{Pavel Jirou\v{s}ek}
\email{jirousek@fzu.cz}
\affiliation{CEICO, Institute of Physics of the Czech Academy of Sciences, Na Slovance 1999/2, 182 21, Prague}
\affiliation{Institute of Theoretical Physics, Faculty of Mathematics and Physics, Charles University, V Hole\v{s}ovi\v{c}k\'{a}ch 2, 180 00 Prague 8, Czech Republic }
\author{Jo\~{a}o Magueijo}
\email{j.magueijo@imperial.ac.uk}
\affiliation{Theoretical Physics Group, The Blackett Laboratory, Imperial College, Prince Consort Rd., London, SW7 2BZ, United Kingdom}
\author{Tom Z\l o\'{s}nik}
\email{zlosnik@fzu.cz}
\affiliation{CEICO, Institute of Physics of the Czech Academy of Sciences, Na Slovance 1999/2, 182 21, Prague}

\date{\today}

\begin{abstract}
In recent work minimal theories allowing the variation of the cosmological constant, $\Lambda$, by means of a balancing 
torsion, have been proposed. It was found that such theories contain {\it parity violating} homogeneous and isotropic solutions, 
due to a torsion structure called the Cartan spiral staircase. Their dynamics are controlled by Euler and Pontryagin {\it quasi}-topological terms in the action. 
Here we show that such theories predict a dramatically different picture for gravitational wave fluctuations 
in the parity violating branch. If the dynamics are ruled solely by the Euler-type term, then linear tensor mode perturbations are entirely undetermined, hinting at a new type of gauge invariance. 
The Pontryagin term not only permits for phenomenologically sounder background solutions (as found in previous literature), but for realistic propagation of gravitational wave modes. We discuss the observational constraints and predictions of these theories. 
\end{abstract}

\maketitle

\section{Introduction}
The cosmological constant, $\Lambda$, and the Copernican principle are two cornerstones of modern cosmology. 
In this paper we explore the implications of the fact that their story may be more intricate than it is usually assumed. 
That the cosmological ``constant''  does not actually need to be constant in theories with torsion has been 
noted, for example, in~\cite{alex1,alex2}. It is not new that torsion can change dramatically the perspective of many problems (for a selection of examples see~\cite{Dolan:2009ni,Mercuri:2009zt,Baekler:2010fr,Poplawski:2011j,Schucker:2011tc,Cid:2017wtf,Zlosnik:2018qvg,Flanagan:2003rb,Sotiriou:2005hu,Bauer:2010jg,Farnsworth:2017wzr,Addazi:2017qus}). It has also been noted~\cite{Baekler:2010fr,MZ} that under the shadow of torsion, homogeneity and isotropy do not imply parity invariance. The Copernican principle therefore has a choice between incorporating parity invariance or not. Parity odd solutions in homogeneous and isotropic models employ a geometrical structure which has been known since the inception of General Relativity: Cartan's spiral staircase~\cite{Baekler:2010fr,spiral}. Thus, a varying Lambda may go hand in hand with parity violating Copernican models, creating an interesting synergy.

Within the theories considered in~\cite{alex1,MZ} the inverse of Lambda 
becomes canonically conjugate to the Chern-Simons invariant\footnote{Models where $\Lambda$ is directly conjugate to the Chern-Simons invariant (or similar quantities) have been considered within the context of unimodular gravity \cite{Henneaux:1989zc,Jirousek:2018ago,Hammer:2020dqp}.}~\cite{Alexander:2018djy,LeeJ}. 
The radical  implications of this fact in quantum cosmology were examined in~\cite{LeeJ} (see~\cite{chopin,Gott:1997pm} for the background problem). 
In the context of classical solutions, the dynamics are then ruled by two topological invariants, of which 
the Chern-Simons functional is the density. Depending on whether one considers the real or imaginary parts of the Chern-Simons term, these are the Pontryagin and  the Euler (or Gauss-Bonnet) invariants. Since these terms appear in the action multiplied
by $\Lambda^{-1}$, they are only topological invariants if $\Lambda$ is 
a constant. The variability of Lambda disrupts their topological nature, and so they are {\it quasi}-topological terms
(to use the terminology of~\cite{alex1}).

The theories considered in~\cite{alex1,alex2,MZ}  have the virtue that they do not add new parameters to gravity with respect to 
Einstein's theory with a cosmological constant. 
The coefficient of the Euler term is fully fixed by the Bianchi identities (from solutions without matter), so that the 
only true new parameter is the numerical coefficient of the Pontryagin term, should we consider it. 
However, for these theories  $\Lambda$ is longer a free parameter, as it is in Einstein's theory. Hence, 
a theory with the Euler term alone  would have 
{\it fewer} free parameters than General Relativity, as argued in~\cite{alex1}. As explained in~\cite{alex2} such a theory conflicts 
dramatically with
basic Hot Big Bang cosmology (it refuses to accept a radiation epoch). The introduction of the Pontryagin term allows for a viable
expansion history (as studied in~\cite{MZ}), leaving the working theory with the {\it same} number of free parameters as General Relativity.

It was found in~\cite{MZ} that the parity-even and parity-odd Copernican solutions belong to separate branches of the dynamics. 
Indeed, a Hamiltonian analysis revealed
 a different structure of constraints and consequently a different number of degrees of freedom. We are therefore talking about 
different phases of the same non-perturbative theory. 
The underlying gauge symmetry associated with the new constraint of the theory is a form of conformal invariance (generalized for theories with torsion). Lambda appears to 
be pure gauge with regards to this symmetry in the parity-even branch (in the absence of matter). The parity-odd branch breaks conformal invariance even in the absence of matter, giving a varying Lambda a  physical meaning. Non-conformal matter does the same in the parity-even branch, but then Lambda becomes a slave to matter (much in the spirit of~\cite{cuscaton}). It is interesting to note that the (odd parity) Pontryagin term is only relevant for the homogeneous and isotropic dynamics
in the  parity-odd branch of the solutions. 

A preliminary investigation~\cite{MZ} revealed that phenomenology in these theories (which, we stress, often have {\it fewer} free parameters than 
General Relativity, and rarely can be made to have more) shows a preference for
 the parity-odd branch in the presence of Pontryagin dynamics. These considerations concerned 
only the background solution, which is already very rich in the parity-odd branch. 
The next obvious step is to investigate the propagation of gravitational waves in the same branch. 
Such is the purpose of the current investigation. 

The plan of this paper is as follows. In Section~\ref{review} we start by reviewing previous results that will be needed in this paper,
translating them into the notation we found most useful for establishing a perturbation calculation. 
In Section~\ref{Sec-pert} we set up the tensor perturbation variables and work out the linearized equations in 
various forms (tetrad index and space-time index forms, and then decomposed in Fourier and helicity modes). 
The equations in general look ominous: we have to contend with first order equations in three variables -- metric, and 
parity even and odd components of the connection --  but in subsection~\ref{Sec-pert} we condense them in a
more aesthetically pleasing form, and lay out a strategy for their solution.  

The rest of the paper is spent on working out solutions for various parameter settings. In Section~\ref{generalfeatures} we briefly discuss general properties of the perturbed equations. Next we discuss a number of limiting cases of interest. As a sanity check we find the 
General Relativity limit in Section~\ref{GRlimit}, with reassuring results. In Section~\ref{Eulertheory} we consider the case where the 
dynamics are ruled purely by an Euler pseudo-topological term.
We unveil our first surprise: the tensor mode perturbation is left entirely undetermined by the equations of motion. This could well signify that they have become a gauge degree of freedom in this case. 

The introduction of the Pontryagin term changes the picture. Physical propagating tensor modes now do exist, but they are endowed with chiral modified dispersion relations. We concentrate on two limiting forms - in Section~\ref{latesol} the propagation of gravitational waves in the late universe is discussed, whilst in Section~\ref{earlytimegravwavepropagation} their propagation at earlier stages when the evolution is dominated by matter and radiation components is discussed. Finally in Section~\ref{outlook} we summarize our results and discuss prospects for further development.

\section{Review of previous results}\label{review}
Here we shall review some results, adapting the notation in previous literature to the notation that shall be more
useful in this paper. Specifically, we shall use the following conventions for indices:
\begin{itemize}
	\item $A,B,C,D$: $SO(1,3)$ gauge indices.
	\item $I,J,K,L$: $SO(3)$ gauge indices.
	\item $\mu,\nu,\alpha,\beta$: spacetime coordinate indices.
	\item $t$: time coordinate index.
	\item $i,j,k,l,$: spatial coordinate indices.
\end{itemize}

\subsection{The full theory and its equations}

The theories we analyze can be written as: 

\begin{widetext}
\be\label{SDcond}
S^g[e,\omega,\Lambda]=-\int \frac{3}{2\Lambda}\left( \epsilon_{ABCD}+\frac{2}{\gamma}\eta_{AC}\eta_{BD}\right)\left(R^{AB}-\frac{\Lambda}{3}e^A e^B\right)
\left(R^{CD}-\frac{\Lambda}{3}e^C e^D\right)-\frac{2}{\gamma}\int T^AT_A.
\ee
\end{widetext}
where $R^{AB} \equiv d\omega^{AB}+ \omega^{A}_{\ph{A}C}\omega^{CB}$, $T^{A} \equiv de^{A}+\omega^{A}_{\ph{A}B}e^{B}$ and unless otherwise stated, multiplication of differential forms is via the wedge product  \footnote{If the parameter $\gamma\rightarrow \infty$ and $\Lambda$ is constrained to be a constant, the resulting theory is the Einstein-Cartan theory alongside an Euler boundary term; the particular coefficient of this boundary term has been found to be associated with interesting properties of Noether charges in gravity \cite{Aros:1999id,Miskovic:2009bm}}.The action can be rewritten as proportional to four terms $S^g=S_{Pal}+S_{Eul}+S_{NY}+ S_{Pont}$, with
\begin{align}
S_{Pal}&=\int  \epsilon_{ABCD}\left( e^A e^B R^{CD} -\frac{\Lambda}{6} e^A e^Be^Ce^D\right)  \label{palatini},\\
S_{Eul}&=-\frac{3}{2}\int \frac{1}{\Lambda} \epsilon_{ABCD}R^{AB}R^{CD} ,\label{eulerterm}\\
S_{NY}&=\frac{2}{\gamma}\int e^Ae^B R_{AB}-T^AT_A ,\\
S_{Pont}&=-\frac{3}{\gamma}\int \frac{1}{\Lambda} R^{AB}R_{AB} .
\end{align}
The first term is the Palatini action, though differs from that of the Einstein-Cartan theory in that we allow $\Lambda$ vary as a dynamical field rather than fixing it to be a constant. The second term is the 
quasi-Euler term of~\cite{alex1}. The third term
is the Nieh-Yan topological invariant (replacing the Holst term should there be torsion). The last term is the quasi-Pontryagin term
studied in~\cite{MZ}. We stress that the connection proposed here between $\gamma$ and the pre-factor of the quasi-Pontryagin term
can be broken, and is not strictly needed. More generally, we 
could also look at theories with arbitrary numerical factors in front of the quasi-Euler and quasi-Pontryagin terms.

As usual, matter can be added to the gravitational action, to yield a total action:
 
\be
S=\frac{1}{2\kappa}S^g(e,\omega,\Lambda)+S_M(\Phi, e,\omega,\Lambda) \label{combined_action},
\ee
where $\Phi$ denote matter fields. The full gravitational equations of motion of this theory are then:
\begin{widetext}
\bea
\epsilon_{ABCD}{\left(e^B  R^{CD}-\frac{1}{3}\Lambda e^B  e^C  e^D\right)}&=&-2 \kappa \tau_A\label{Eeq}\\
T^{[A}  e^{B]} +\frac{3}{2\Lambda^2}d\Lambda  R^{AB}-\frac{3}{4\gamma \Lambda^2}\epsilon^{ABCD}d\Lambda R_{CD} &=& \kappa {\cal S}^{AB}\label{omEq}\\
\epsilon_{ABCD}{\left(R^{AB}  R^{CD}-\frac{1}{9}\Lambda^2 e^A   e^B  e^C  e^D\right)}+\frac{2}{\gamma}R^{AB}R_{AB}&=& -2\kappa {\cal J}\label{Leq}
\eea
\end{widetext}
where $\kappa \equiv 8\pi G$ and we have defined energy momentum 3-form $\tau_A =\frac{1}{2}\frac{\delta S_M}{\delta e^A}$, the spin-current 3-form ${\cal S}^{AB} \equiv -(1/2)\eps^{ABCD}\delta S_{M}/\delta \omega^{CD}$ and the $\Lambda$-source 4-form ${\cal J} \equiv (2/3)\delta S/\delta \Lambda$. They are obtained by varying  (\ref{SDcond}) together with the action for matter with respect to $e$, $\omega$ and $\Lambda$, respectively. A key property of these models is that Einstein's equation (\ref{Eeq}) takes the same form in the Einstein-Cartan formulation of gravity (where $\Lambda=cst.$). Any dynamics for $\Lambda$ will arise from the gravitational field $\omega^{AB}$ rather than via the addition of explicit kinetic terms for $\Lambda$ in the Lagrangian. 

In this paper we will confine ourselves to situations where the quantities ${\cal S}^{AB}$ and ${\cal J}$ both are negligible. For standard `minimal' coupling between fermions and the spin connection, the quantity ${\cal S}^{AB}$ is sourced by the axial spinor current; much of our focus will be on the behaviour of certain cosmological perturbations in `recent' post-recombination cosmological history where this quantity is expected to be negligible \cite{Dolan:2009ni}. The assumption that ${\cal J}$ is negligible must be regarded as a simplifying assumption and more detailed analysis is needed to determine its expected coupling to matter. For the particular cosmological consequences of the theory examined in this paper, it will suffice to that the matter content is describable in terms of perfect fluids. By way of example, a perfect fluid with density $\rho$, pressure $p$ and four-velocity $U^{\mu} = e^{\mu}_{A}U^{A}$ will have stress-energy 3-form:

\begin{align}
\tau^{A} =-\frac{1}{6} (\rho+ p) U^{A}\eps_{BCDE}U^{B}e^{C}e^{D}e^{E}
-\frac{1}{6}p \eps^{A}_{\ph{A}BCD}e^{B}e^{C}e^{D}
\end{align}

\subsection{The background solution}
We now look at the behaviour of the theory in situations where spacetime has Friedmann-Robertson-Walker (FRW) symmetry. This symmetry 
is widely considered to well approximate the geometry of the universe on large scales and there exist strong constraints on the evolution of the universe within this framework. We will henceforth refer to possible solutions with this symmetry as `background' solutions as later we will consider the behaviour of small perturbations around them.  It is important then to demonstrate that the combined action (\ref{combined_action}) yields solutions that are consistent with these constraints.

We shall denote all background quantities by a bar over the respective variable. 
For simplicity we assume that the background spatial curvature is zero, so that we can use Cartesian 
coordinates with 
\bea 
\bar  e^0&=&N(t)dt\\
\bare^{I} &=& a(t)\delta^{I}_{i} dx^{i}
\eea
where $N(t)$ is the lapse function ($N=1$ for proper time) and $a(t)$ is the expansion factor. 
Note that $\bar{e}^{I}_{i}= a\delta^{I}_{i}$ and 
$\bar{e}^{i}_{I} = a^{-1}\delta^{i}_{I}$. 
Then, the spin connection will be given by:
\begin{align}
\bar{\omega}^{0I} &=  g(t)a(t) \delta^{I}_{i }dx^{i} \label{woiback} \\
\bar{\omega}^{IJ} &= -P(t)a(t) \eps^{IJ}_{\phantom{IJ}K}\delta_{k}^{K} dx^{k} \label{wijback}
\end{align}
where $g$ and $P$ are its parity even and odd components, respectively. A connection of the form (\ref{wijback}) was considered by Cartan as an extension to Riemannian geometry, with parallel transport according to this connection yielding a rotation of vectors with a `handedness' dictated by the sign of $P$. This effect has been termed Cartan's spiral staircase and we will see that all parity violating effects in this gravitational model appear only when $P\neq 0$. The torsion associated with (\ref{woiback}) and (\ref{wijback}) is given by:
\bea
\bar{T}^{0} &=&   0\\
\bar{T}^{I} &=&T \bar{e}^{I}\bar{e}^{0}+P\eps^{IJK}\bar{e}_{J}\bar{e}_{K} 
\eea
with the parity even component $T$ related to $g$ by:
\be
T=\left(g-\frac{1}{N}\frac{\dot{a}}{a}\right).
\ee
The field strength is:
\bea
\bar R^{0I }&=&  \frac{1}{N}\left(\dot{g}+\frac{\dot{a}}{a}g\right)\bare ^{0}\bare ^{I} + gP  \eps^{IJK}\bare _{J}\bare _{K} \\
\bar R^{IJ} &=& \frac{1}{N}\left(\dot{P}
+P\frac{\dot{a}}{a}\right) \eps^{IJ}_{\ph{IJ}K}\bare ^{K}\bare ^{0}+\left(g^{2}-P^{2}\right)\bare ^{I}\bare ^{J}.
\eea

It can be shown that with this ``Copernican'' ansatz, equations (\ref{Eeq}) to (\ref{Leq}) become:
\begin{widetext}
\bea
g^{2}-  P^{2} &=&\frac{\Lambda+\kappa\rho }{3}\label{EeqF}\\
\frac{(ag)^.}{a}&=& \frac{\Lambda}{3}-\frac{\kappa}{6}(\rho +3p)\label{ray}\\
T&=&\frac{\dot\Lambda}{2\Lambda^2}\left(\Lambda+ \kappa\rho -\frac{6}{\gamma}gP\right)\label{TeqF}\\
P&=&\frac{3\dot{\Lambda}}{\Lambda^2}\left(gP +\frac{\Lambda +\kappa\rho}{6\gamma}\right)\label{PeqF}\\
(\Lambda+\kappa\rho)
\left(\Lambda-\frac{\kappa}{2}(\rho +3p)\right)-\Lambda^2&=&18g
 P\frac{(a P)^.}{a} 
+\frac{9}{\gamma}\left(\frac{\Lambda+\kappa\rho}{3}\frac{(aP)^.}{a} +\frac{2}{3}\left(\Lambda-\kappa\frac{\rho+3p}{2}\right)gP\right)
\label{LeqF}
\eea
\end{widetext}
As shown in Appendix \ref{backeqs}, this system can be cast in the form of a first-order system of evolution equations for $\{a,g,\Lambda,P\}$ plus a constraint (the Hamiltonian constraint/Friedmann's equation). Reference to these background equations will be made at several points in this paper, 
to simplify the perturbation equations. 

\subsection{Background evolution}
\label{backgroundevolution}
We now discuss solutions to equations (\ref{EeqF})-(\ref{LeqF}) with an emphasis on solutions that appear likely to be most consistent with the observed expansion history of the universe. Care must be taken here as many probes of background quantities are additionally sensitive to details of cosmological perturbations. For example, the position of the first peak of temperature anisotropies in the cosmic microwave background (CMB) is sensitive to both the distance to last scattering (a background quantity) and the sound horizon at last scattering (a quantity which additionally depends on the form of equations describing cosmological perturbations) \cite{Ma:1995ey}.

The system of equations (\ref{EeqF})-(\ref{LeqF}) is rather complicated and must be solved numerically. However, relevant approximate solutions do exist, which we will now discuss.

\subsubsection{Early times}
\label{earlytime}
There is strong evidence that the universe has undergone an early period (`the radiation era') where relativistic species (such as photons and relativistic neutrinos) dominate the evolution of the universe for a time before the universe cools down enough such that the gravitational effect of near-pressureless/dustlike matter (baryons and dark matter) dominates (`the matter era'), before eventually a new source of energy - typically termed dark energy - begins to dominate and cause the expansion of the universe to accelerate \cite{PlanckCollaboration2015}. We will look to see whether the theory (\ref{combined_action}) permits this kind of cosmological history, whilst ascribing the recent cosmological acceleration to - now dynamical - $\Lambda$. An important part of this is that the gravitational effect of new degrees of freedom quantities such as $\Lambda$ and the torsion $P$ do not contradict the above picture.

It can be shown that when $|\gamma| \ll 1$, to first order in $\gamma$ there exists a solution for the field $P$ in the limit $\Lambda \rightarrow 0$

\begin{align}
P = P_{(\rho)} =  \frac{\gamma}{3}\sqrt{\frac{\kappa\rho}{3}} \label{ptrack}
\end{align}
We see then that when this solution holds, the torsion field $P$ is proportional to $\gamma$ and so a smaller value of $\gamma$ suppresses torsion in the cosmological background. Neglecting the contribution of $\Lambda$ is expected to be a good approximation in the earlier universe where the `dark energy' is a sub-dominant contributor to the universe's expansion. When (\ref{ptrack}) holds it may be shown that the Friedmann equation can be recovered in approximation:

\begin{align}
	3 \bigg(\frac{\dot{a}}{a}\bigg)^{2} &= \bigg(1-\frac{\gamma^{2}}{9}\bigg)\kappa\rho + {\cal O}(\gamma^{3})
	\end{align} 
where we have adopted the $N=1$ spacetime gauge. Hence the solution (\ref{ptrack}) acts to rescale the bare Newton's constant $G$
during times when the effect of $\Lambda$ is negligible. The degree to which this effect is observable depends on how the value of Newton's constant $G_{N}$ measured in tabletop experiments is related to $G$. If $G\neq G_{N}$, then the rate of expansion $\dot{a}/a$ due to a given $\rho$ will be different from as is the case in General Relativity and so in principle $\gamma$ could be constrained by probes of the expansion rate during big bang nucleosynthesis \cite{Carroll:2004ai}.

However, importantly, the solution (\ref{ptrack}) is not stable. By way of illustration, we  may consider the evolution of small, homogeneous perturbations $P= P_{(\rho)}(1+\delta_{P}(t))$. It can be shown that deep in the radiation era where $\kappa\rho/3 \sim H_{0}^{2}\Omega_{r}/a^{4}$ - where $H_{0}$ is the Hubble constant today - that 

\begin{align}
	P_{(\rho)} &= \frac{\gamma}{3a^{2}}H_{0}\sqrt{\Omega_{r}} \\
	\delta_P &= {\cal C} a^{3}
\end{align}
Therefore $\delta_{P}$ grows as $a$ increases. By way of example, if $\delta_{P} \ll 1$ at $a \sim 10^{-15}$ then for it to remain smaller than unity at $a \sim 10^{-5}$ we must have $\delta_{P}(a=10^{-15}) < 10^{-30}$. This indicates that significant fine-tuning of initial data is required for the spiral staircase field $P$ to find itself following the solution (\ref{ptrack}).

If $P$ deviates considerably from the tracking solution, the tendency is for $P$ to evolve to \emph{dominate} the evolution of the universe. In this case it may be shown that $P = P_{0}/a$ where $P_{0}$ is a constant and $a \sim (1+\gamma)(t-t_{0})$ - here the evolution of the universe due to $P$ resembles a General-Relativistic \emph{empty} universe with negative spatial curvature. It is hard to see how such a universe could be consistent with experiment. This is the case even if $P$ is initially negligible. Therefore, the phenomenological viability of the model rests on $P$ being able to 
initially find itself sufficiently close to the form (\ref{ptrack}) to avoid dominating the evolution of the universe.

\subsubsection{Late times}
\label{latetime}
During late time cosmological evolution for realistic cosmologies we expect that universe to begin accelerating and we look to ascribe this to $\Lambda$ and $P$ beginning to dominate cosmic evolution. Again assuming $|\gamma|\ll 1$ and now assuming $P^{2} \ll \Lambda$ and taking the limit $\rho\rightarrow 0$ we have the following evolution equations for $\Lambda$ and $P$:

\begin{align}
\frac{dP}{d \ln a} = -3P, \quad \frac{d\Lambda}{d\ln a} &= 2\sqrt{3}\gamma P \sqrt{\Lambda}
\end{align}
which possess solutions

\begin{align}
P &= \frac{P_{i}}{a^{3}} \label{latep}\\
\Lambda &= \Lambda_{0}-\frac{2\gamma}{\sqrt{3}a^{3}}P_{i}\sqrt{\Lambda_{0}}
\end{align}
So, asymptotically for large $a$, $\Lambda \rightarrow \Lambda_{0}$ and $P\rightarrow 0$, leading to a confluence with the current standard cosmological picture of the late-time universe's evolution being dominated by a cosmological constant of magnitude $\Lambda_{0}$. The contribution of $P$ to the Hamiltonian constraint goes as $\sim P^{2}$ so we see that in this regime $P$ evolves like a shear component, its energy density diluting as $a^{-6}$. For realistic cosmologies a typical value for $P_{i}$ will be given by its value when $\Lambda$ begins to dominate the evolution of the universe at a scale factor $a\sim a_{i}$ following a period of matter domination during which $P \sim (\gamma/3)H_{0}a_{i}^{-3/2}$ (from the solution (\ref{ptrack})). We expect then $P_{i} \sim (\gamma/3)H_{0}a_{i}^{3/2}$. This has important implications: if phenomenologically viable cosmologies involve $P$ staying on the tracking solution (\ref{ptrack}) for an appreciable amount of time, this means that fixing $\gamma$ fixes the size of $P$ during matter domination, and the size of $P_{i}$ as the cosmological constant begins to dominate.

\begin{figure}
	\center
	\epsfig{file=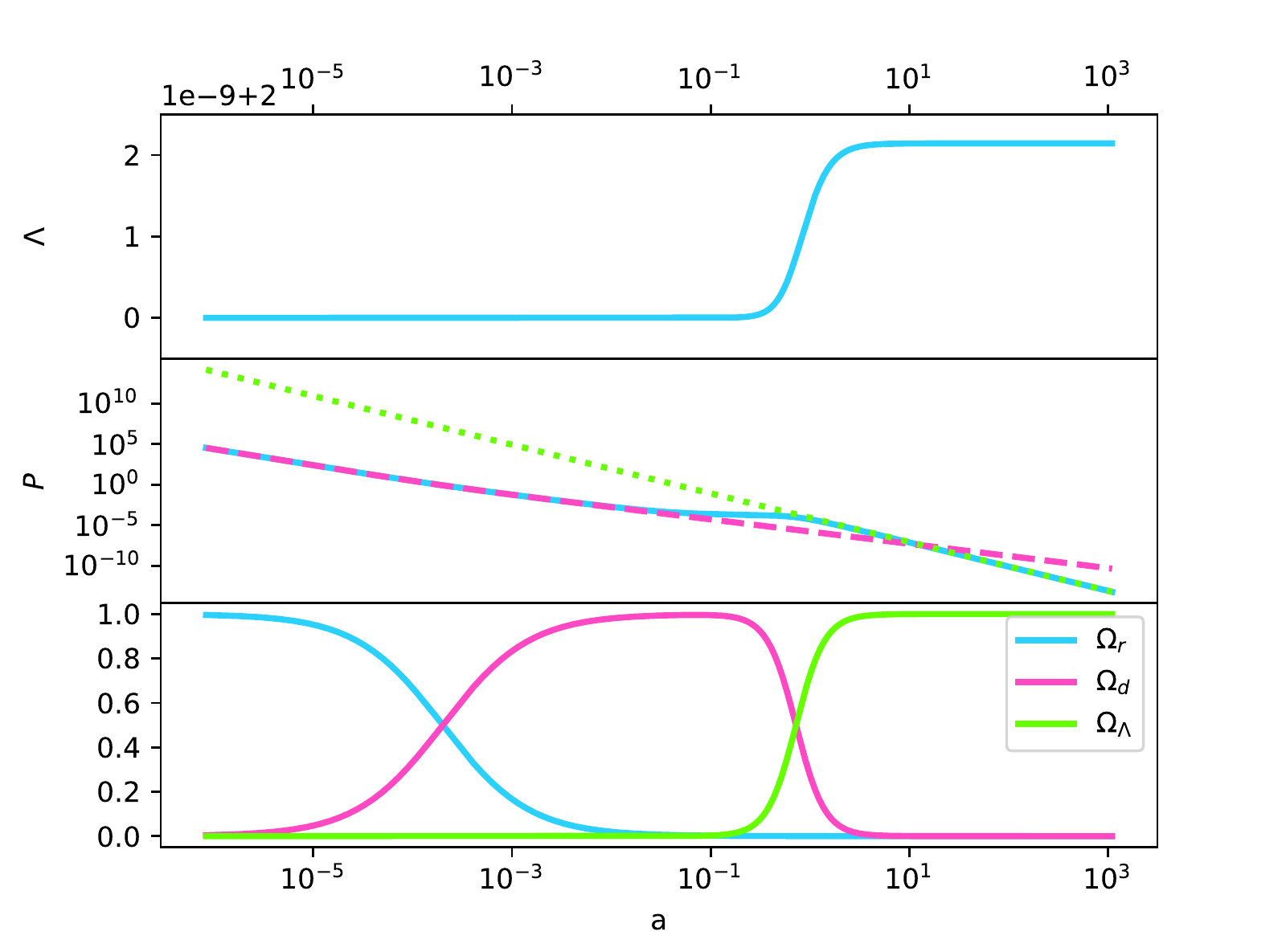,width=9.cm}
	\caption{The numerical evolution of various quantities for the parameter choice $\gamma = 10^{-5}$. In the upper plot the evolution of $\Lambda$ is shown; it can be seen that the field changes by roughly one part in $10^{9}$ over cosmic history. In the middle plot the solid line shows the exact evolution of the torsion field $P$ whilst dashed line shows the solution (\ref{ptrack}) and the dotted line shows the solution (\ref{latep}). 
		The lower plot shows evolution of $\Omega$ quantities (here defined as fractional contributions to $g^{2}$ in the Hamiltonian constraint) and $P$ (solid lines) as a function of $\ln a$ for a realistic universe. Subscripts $d$ and $r$ denote dust and radiation-like components of the universe. The scale factor is fixed to be $a=1$ at the present moment, and units where the present day Hubble parameter $H_{0}$ is set to unity are used.}
	\label{figure1}
\end{figure}

We now discuss the evolution of $\Lambda$. It can be seen from (\ref{ptrack}) that the tracking solution can exist if $\mathrm{sign}(P) = \mathrm{sign}(\gamma)$. Recall that the $\Lambda$ equation of motion is $\dot{\Lambda} = 2 \gamma  \Lambda ^2 P/(6 \gamma  g P+\Lambda+\kappa\rho)$ and therefore in the earlier universe if $\kappa\rho$ is initially greater than $\Lambda$ it will tend to suppress time variation of $\Lambda$. Furthermore, we will have $\dot{\Lambda} > 0$ throughout, meaning that $\Lambda$ must be of smaller magnitude in the past than today. A typical evolution of $\Lambda$ and $P$ are shown in Figure \ref{figure1}.

In summary then, numerical exploration suggests that unless $P$ finds itself on the tracking solution (\ref{ptrack}) for much of cosmic history, it will tend to dominate the evolution of the universe and therefore very likely in conflict with cosmological data. This requires fine tuning of the initial value of $P$ so that it begins close to the tracking solution. During the tracking stage, the effect of $P$ is to rescale Newton's constant $G$. We will find later that deviations of gravitational wave speed from unity tend to be of order $\gamma^{2}$. This justifies our assumption that $|\gamma| \ll 1$ and - in conjunction with recent constraints on the speed of gravity -  restricts the fractional rescaling of $G$ to be ${\cal O}(10^{-15})$, which is well within bounds that will be placed by BBN constraints for the foreseeable future \cite{Carroll:2004ai}. Additionally we see that a smaller value of $\gamma$ tends to lower the total time variation of $\Lambda$ over cosmic history, making it more difficult to distinguish from a genuine cosmological constant.

\section{The perturbed equations of motion for tensor modes}\label{Sec-pert}
We now look at the evolution of small perturbations to the cosmological background. We perturb the tetrad and connection as:
\begin{align}
\delta e^{0} &=  0\\
\delta e^{I} &=   \frac{1}{2}{\cal H}^{IJ}\bare_{J}\\
\delta \omega^{0I} &=  \frac{1}{2}{\cal E}^{IJ}\bar{e}_{J}\label{taudef}\\
\delta \omega^{IJ} &=   \frac{1}{2}\eps^{IJ}_{\ph{IJ}K}{\cal B}^{KL}\bar{e}_{L}\label{chidef}
\end{align}
where ${\cal H}^{[IJ]}={\cal E}^{[IJ]}={\cal B}^{[IJ]}=0$ and ${\cal H}^{I}_{\ph{I}I}={\cal E}^{I}_{\ph{I}I}={\cal B}^{I}_{\ph{I}I}=0$.
In addition we apply the restriction of looking at tensor (transverse, traceless) modes, so that we impose:
\be
\bd_{I}{\cal H}^{IJ} = \bd_{I}{\cal E}^{IJ} = \bd_{I}{\cal B}^{IJ} =0. \label{vanishingdivergence}
\ee
where $\bar{D}_{I} \equiv \bar{e}^{i}_{I}\bar{D}_{i}$ and $\bar{D}_{i}$ is the covariant derivative according to $\bar{\omega}^{IJ}_{\ph{IJ}i}$. Note the field $P$ does not contribute to the expressions (\ref{vanishingdivergence}) and so the equations are equivalent to $\bar{e}^{i}_{I}\partial_{i}{\cal H}^{IJ}= \bar{e}^{i}_{I}\partial_{i}{\cal E}^{IJ}= \bar{e}^{i}_{I}\partial_{i}{\cal B}^{IJ}=0$.

Given a quantity $Y_{IJ}$ that represents a small perturbation, it can be converted into a tensor $Y_{ij}$ in the spatial coordinate basis via $Y_{ij} \equiv \bar{e}^{I}_{i}\bar{e}^{J}_{j}Y_{IJ}$. Given our assumption of vanishing spatial curvature, a `co-moving' tensor $\tilde{Y}_{ij} = Y_{ij}/a^{2}$ can further be constructed.

The linearly-perturbed form of equations (\ref{Eeq})-(\ref{Leq}) can be written as a system of linear partial differential equations. For simplicity we decompose these perturbations into plane-wave Fourier  components labelled by wave vector $k^{i}$  and as a further simplification we decompose each co-moving tensor mode Fourier mode into helicity eigenstates i.e.:

\begin{align}
\tilde{Y}_{ij}({\vec{x}},t) &=\frac{1}{(2\pi)^{3}} \int d^{3}k\sum_{\pm}Y^{\pm}(\vec{k},t)e^{ik_{i}x^{i}} \tilde{\cal P}^{\pm}_{ij} 
\end{align}
Here  $\tilde{\cal P}^{\pm}_{ij}$ are co-moving polarization tensors for $+$ and $-$ helicity components. Without loss of generality we will focus the remainder of the analysis on modes that propagate along the z-direction of the coordinate system i.e. $\vec{k}=(0,0,k)$. We then have the following important identities:

$
i k^{m}\tilde{\eps}_{im}^{\ph{im}l}\tilde{\cal P}^{\pm}_{lj} = \pm k \tilde{\cal P}^{\pm}_{ij}
$
and
$\tilde{\cal P}^{\lambda}_{ij}\tilde{\cal P}^{\lambda' ij} = 2\delta^{\lambda\lambda'}$  where $\lambda = +,-$ and $\tilde{\eps}_{ijk}$ is the co-moving three dimensional Levi-Civita symbol. Indices of co-moving tensors are taken to be raised and lowered with the Kronecker delta symbol.

After some algebra, it can be shown that the spin connection equations of motion yield the following equations:

\begin{align}
\bigg(\begin{array}{c}{\cal B}^{\pm} \\
{\cal E}^{\pm} \end{array}\bigg) &=\frac{1}{A^{2}+B^{2}}\bigg( \begin{array}{cc}
A & B \\
-B & A \end{array} \bigg)\bigg(\begin{array}{c} -k_{P}^{\pm}{\cal H}^{\pm} \\
\dot{\cal H}^{\pm}+\bigg(2\frac{\dot{a}}{a}-g\bigg){\cal H}^{\pm} \end{array}\bigg)\label{matrixBE}
\end{align}
where here and subsequently we choose the spacetime gauge $N=1$ (proper time) and where 
\begin{align}
A= & 1-\frac{3\dot{\Lambda}}{\Lambda^{2}}\bigg(g+k_{P}^{\pm}\gamma^{-1}\bigg)\label{Aeq}\\
B= &  \frac{3\dot{\Lambda}}{\Lambda^{2}}\bigg(k_{P}^{\pm} - g\gamma^{-1}\bigg) \label{Beq}
\end{align}
We have introduced the polarization-dependent,  torsion-adjusted proper wavenumber $k_{P}^{\pm}$ according to:

\begin{align}
k_{P}^{\pm} \equiv \pm \frac{k}{a}-P
\end{align}
For reference, in the usual Einstein-Cartan theory we have $A=1$, $B=0$; in that case, ${\cal B}^{\pm}$ is related to spatial derivatives of ${\cal H}^{\pm}$ and ${\cal E}^{\pm}$ is related to time variations of ${\cal H}^{\pm}$. All modifications to the relation between $\{{\cal E}^{\pm},{\cal B}^{\pm}\}$ and ${\cal H}^{\pm}$ stem from non-constancy of $\Lambda$. Hence  the connection equations imply
that in general the parity even and odd components  of the connection (${\cal E}$ and ${\cal B}$) can be obtained from their
Einstein-Cartan expressions via a rotation, with an angle $\theta$ satisfying:
\be
\tan\theta=\frac{B}{A}=\frac{ \frac{3\dot{\Lambda}}{\Lambda^{2}}(g-\gamma k_{P}^{\pm})\gamma^{-1}}{1- \frac{3\dot{\Lambda}}{\Lambda^{2}}(g+k_{P}^{\pm}\gamma^{-1})}
\ee
followed by a dilatation by $1/\sqrt{A^2+B^2}$. 

Then we may look to
find the ``second order''  evolution equation for ${\cal H}^{\pm}$ by inserting the solution (\ref{matrixBE}) for ${\cal B}^{\pm}$ and ${\cal E}^{\pm}$ into the Einstein equation:
\begin{align}\label{EeqP}
0 &= \dot{\cal E}^{\pm}+\bigg(\frac{\dot{a}}{a}+g\bigg){\cal E}^{\pm}-k_{P}^{\pm}{\cal B}^{\pm} \nn\\
&-\bigg(\frac{2}{3}\Lambda+\frac{\kappa}{6}(\rho-3p)\bigg) {\cal H}^{\pm}
\end{align}
We now look at solutions to the system (\ref{matrixBE}) and (\ref{EeqP}).

\section{General features}
\label{generalfeatures}
Generally, if the solution (\ref{matrixBE}) is inserted into (\ref{EeqP}) then the resulting coefficient of $\ddot{\cal H}^{\pm}$ is proportional to:

\begin{align}
\bigg(1-\frac{6P}{\Lambda+\kappa\rho}k_{P}^{\pm}+ \dots\bigg). 
\end{align}
Where dotes denote terms of higher order in $|\gamma| \ll 1$. We see that the coefficient is not positive-definite and hits zero when $k= k^{\pm}_{*}$:
\begin{align}
k^{\pm}_{*} =  \pm \frac{a}{6P}\bigg(\Lambda+\kappa\rho+6P^{2}\bigg) \label{totalcollapse}
\end{align}
For example in the very late universe we may expect $\Lambda \sim \Lambda_{0} = cst.$ to dominate the evolution of the universe hence then:
\begin{align}
k^{\pm}_{*} \sim  \pm \frac{a}{6P}\Lambda_{0} \label{kstar}
\end{align}
where we've assumed that $P^{2}/\Lambda_{0} \ll 1$. 

Following the arguments proposed in Subsections \ref{earlytime} and \ref{latetime}, we have that $k_{*}^{\pm} \sim \pm 2a^{4}\Lambda_{0}/\gamma$ for realistic cosmologies.
Reaching $k^{\pm}_{*}$ likely corresponds to the coefficient of the kinetic term of one of the gravity wave polarizations in the perturbed Hamiltonian passing through zero, signalling that the mode is becoming ghostlike. If the theory is to be experimentally viable then the scales $k_{*}^{\pm}$ should be of far greater magnitude than those for which the approximation of gravitational waves as small perturbations obeying linear equations of motion is expected to hold well. We will see that recent observations from the LIGO experiment severely constrain the value of $|\gamma|$ because of this.

As for the case of $P(t)$, we see that a key parameter for the size of $k^{\pm}_{*}$ is $\gamma$. 

When $k\neq k_{*}^{\pm}$ and with the important exception of the limit $|\gamma| \rightarrow \infty$  (see Section \ref{Eulertheory}), it is possible to write the Einstein equation (\ref{EeqP}) in the following form :

\begin{align}
	\ddot{\cal H}^{\pm} &= -\omega_{\pm}^{2}(k,t){\cal H}^{\pm}-f_{\pm}(k,t)	\dot{\cal H}^{\pm} \label{eingen}
\end{align}
For arbitrary values of $\gamma$, the form of $\omega_{\pm}^{2}(k,t)$ and $f_{\pm}(k,t)$ will be extremely complicated and so we will concentrate in detail on how (\ref{eingen}) looks in relevant, limiting cases.

\section{The Einstein-Cartan limit}\label{GRlimit}
We start by finding the Einstein-Cartan 
limit of these theories, noting that when $\gamma$ is finite and $\rho=p=0$ there are solutions where $P=0$, $g=\dot{a}/a$ and
$\Lambda$ is constant~\cite{MZ}. Taking these background solutions we should obtain the Einstein-Cartan limit for our theory,
which is equivalent to General Relativity in this situation. 
Inserting these conditions into the formalism just developed, we find $\dot{\Lambda}=0$, and so $A=1$ and $B=0$,
as already announced in the previous Section. 
The connection equations are therefore:
\begin{align}
{\cal B}^{\pm} &= - k_{P}^{\pm}   {\cal H}^{\pm}\label{GRchi}\\
{\cal E}^{\pm}  &=  \bigg(\frac{d}{dt}  +\frac{\dot{a}}{a}\bigg) {\cal H}^{\pm} .\label{GRtau}
\end{align}
Note that since $P=0$ we have $k_{P}^{\pm}=\pm k/a$, and so for gravity waves in Einstein-Cartan theory, the parity-odd connection perturbation, $\cal B$, is a spatial gradient of the metric,
whereas the parity-even component, ${\cal E}$, is a time derivative of the metric (cf. Eqs~(\ref{chidef}) and (\ref{taudef})). Inserting these expressions into the Einstein equation (\ref{EeqP}), as prescribed, 
we find:
\be
\ddot {\cal H}^{\pm} +3\frac{\dot{a}}{a}\dot {\cal H}^{\pm} + \left(\dot g + 2g^2 -\frac{2}{3}\Lambda\right){\cal H}^{\pm} + (k_{P}^{\pm})^2{\cal H}^{\pm}=0\label{GRproptoeq}
\ee
where the dot denotes derivative with respect to the background proper time. In the Einstein-Cartan theory we have $T=0$ in the absence of background sources of torsion, so
$g=\dot a/a$, and the background equations of motion read (see (\ref{EeqF}) and (\ref{ray})):
\bea
g^2&=&\frac{\Lambda}{3}\\
\dot g+\frac{\dot a}{a}g=\dot g+ g^2&=&\frac{\Lambda}{3}. 
\eea
Therefore we find:
\begin{align}
\ddot {\cal H}^{\pm} +3\frac{\dot a}{a}\dot {\cal H}^{\pm}&=  - (k_{P}^{\pm})^2 {\cal H}^{\pm} \nn\\
&= -\bigg(\frac{k}{a}\bigg)^{2}{\cal H}^{\pm} \label{grgw}
\end{align}
Thus, our formalism for gravity waves reduces to the textbook equations for gravity waves in General Relativity in this limit.

\section{Euler theory ($\gamma\rightarrow \infty$) in a parity-odd background ($P\neq 0$)}
\label{Eulertheory}

Our first surprise arises when we consider a theory with the Euler pseudo-topological term only, by letting
$\gamma \rightarrow \infty$, but with a background with $P\neq 0$. 
Then, as the background Equation~(\ref{PeqF}) shows (with $P\neq 0$ and 
$\gamma\rightarrow \infty$), we must have $3\dot{\Lambda}g=\Lambda^{2}$. 
Therefore, the definitions of $A$ and $B$  
(Eqns.~(\ref{Aeq}) and  (\ref{Beq})) lead to:
\bea
A&=&0, \quad  B= \frac{k_{P}^{\pm}}{g}
\eea
These are orthogonal to the Einstein-Cartan values, in the sense that  
for the latter the matrix (\ref{matrixBE}) is diagonal, whereas here the matrix is purely off-diagonal.  
Indeed the rotation part of the transformation is now  $\theta=\pi/2$. 
This is reflected in the way the connection is related to the metric. 
the Einstein-Cartan case (cf. Eqns. (\ref{GRchi}) and (\ref{GRtau})) we have:
\begin{align}
{\cal E}^{\pm} &=  g {\cal H}^{\pm}\\
k_{P}^{\pm}{\cal B}^{\pm}&=
g \bigg(\dot{\cal H}^{\pm} +\bigg(\frac{2\dot{a}}{a}-g\bigg){\cal H}^{\pm} \bigg)
\end{align}
Inserting into the Einstein equation (\ref{EeqP}) we find that
not only  does this imply an absence of second order time derivatives for ${\cal H}^{\pm}$, but the first time derivatives cancel out.
In addition the algebraic equation obtained is 
\begin{align}
 \bigg(\frac{1}{2}\dot{g}+g^{2}-\frac{g}{2}\frac{\dot{a}}{a}-\frac{\Lambda}{3}+\frac{1}{12}(3p-\rho)\bigg){\cal H}^{\pm} =0 \label{imundetermined}
\end{align}
The term multiplying ${\cal H}^{\pm}$ in (\ref{imundetermined}) vanishes due to the background equations of motion, therefore the tensor mode perturbation is ${\cal H}^{\pm}$  completely undetermined by the perturbed equations of motion \footnote{This would appear to contradict the result found in \cite{Barrientos:2019awg} which says that tensor modes propagate luminally as in General Relativity in a model with tensor mode perturbation equations that should be mappable to the ones considered here.}.

One may wonder to what extent this is a result of the particular choice for our action. For example, if the coefficient $-\frac{3}{2}$ in the term (\ref{eulerterm}) is replaced by $-\frac{3}{2\xi}$ then it can be shown that Einstein's equation instead becomes:

\begin{align}
 \frac{1}{\xi}\bigg(1-\xi\bigg)\bigg(4\Lambda+\rho-3p\bigg){\cal H}^{\pm} =0 \label{imdeterminedbutzero}
\end{align}
Thus in the case when $\Lambda\neq 0$, $\rho\neq 3p$ and $\xi\neq 1$, the perturbation $\mathcal{H}^{\pm}$ is not undetermined but fixed to vanish. It appears that the presence of the Euler term in the absence of the Pontryagin term is sufficient to nullify the dynamics of the perturbation $\mathcal{H}^{\pm}$ with the existence of a special case $\xi=1$, which leaves them undetermined by the perturbed equations of motion. Note that the case of simultaneous vanishing of the Euler and Pontryagin term ($\xi\rightarrow\infty$) does not correspond to General Relativity. In fact such limit yields a rather exotic background solution $a=0$ due to $\Lambda$ being a dynamical field.

\section{The leading order solution for the general case in the late universe}
\label{latesol}

We now consider a general finite value of $\gamma$ and look at the perturbed equations in a regime where the evolution of the universe is dominated by $\Lambda$. We define a dimensionless parameter $\epsilon_{P} \equiv P/\sqrt{\Lambda}$ which is expected to be of magnitude much smaller than unity in the late universe. Furthermore we  assume that $|\gamma| \ll 1$.
Inserting the solutions for ${\cal E}^{\pm}$ and ${\cal B}^{\pm}$ from (\ref{matrixBE}) into the Einstein equations and keeping only terms up to second order in $\{\epsilon_{P},\gamma\}$ we find:

\begin{widetext}
	\begin{align}\label{mastereq}
	\ddot{\cal H}^{\pm} &= -\omega_{\pm}^{2}(k,t){\cal H}^{\pm}-f_{\pm}(k,t)	\dot{\cal H}^{\pm} \\
	\omega^{2}_{\pm}(k,t) &\equiv \bigg[ 1-\kappa\frac{\big(\rho(-\Lambda+\kappa\rho)+3p(\Lambda+\kappa\rho)\big)}{(\Lambda+\kappa\rho)^{2}}\gamma ^2 \nn\\
	& - \frac{8 \sqrt{3} \left( \Lambda ^{5/2}+ \kappa  \Lambda
		^{3/2} \rho- \kappa ^2 \sqrt{\Lambda } \rho ^2 -3  \kappa ^2 \sqrt{\Lambda } p \rho -3  \kappa 
		\Lambda ^{3/2} p\right)}{(\kappa  \rho +\Lambda )^{5/2}} \gamma\eps_{P}+\frac{42  \left(2 \Lambda ^2-\kappa  \Lambda  \rho -3 \kappa  \Lambda  p\right)}{(\kappa  \rho
		+\Lambda )^2}\epsilon_{P}^{2}\bigg]\bigg(\frac{k}{a}\bigg) ^{2}\nn\\
	&	+ {\cal O}(\epsilon_{P},\gamma)^{3}\\
	 f_{\pm}(k,t)& \equiv \sqrt{3(\Lambda+\kappa\rho)} + \bigg[\mp\gamma\kappa\frac{1}{(\Lambda+\kappa\rho)^{2}}(\rho(-\Lambda+\kappa\rho)+3p(\Lambda+\kappa\rho)) \pm 4\sqrt{3} \eps_{P}\sqrt{\Lambda}\frac{1}{(\Lambda+\kappa\rho)^{3/2}}(3p\kappa-2\Lambda+\kappa\rho)\bigg]\bigg(\frac{k}{a}\bigg) \nn\\
 	& + {\cal O}(\epsilon_{P},\gamma)^{2}
	\end{align}
\end{widetext}
Roughly speaking, positivity of both $\omega^{2}_{\pm}(k,t)$ and $f_{\pm}(k,t)$ imply that ${\cal H}^{\pm}$ evolves in a stable manner.

Following \cite{Johan} (see also~\cite{Creminelli_2017,Sakstein_2017,Ezquiaga_2017}) the speed of monochromatic tensor modes today $c^{\pm}_{T}$ (taking $a=1$) is given by $c_{T}^{\pm} = \frac{\omega_{\pm}}{k}$. In general our expression for $c_{T}$ will be rather complicated but it is instructive to detail the order of magnitude of terms appearing in its expressions.  Given how we expect $P(t)$ to scale with $\gamma$ from the results of Section \ref{latetime} we find that:

\begin{align}
c_{T}^{\pm} &\sim 1 + {\cal O}\bigg(\gamma^{2}\bigg) + \dots \label{ctpm}
\end{align}

Constraints from the LIGO experiment roughly constrain the deviation of $c_{T}^{\pm}$ from unity by approximately $10^{-15}$. The constraint on the speed of gravitational wave speed then places the following restriction on $\gamma$:

\begin{align}
\gamma^{2} \lesssim {\cal O}(10^{-15})
\end{align}
Given this constraint and the small value of $H_{0}/k_{LIGO}$, the remaining immediate constraint from $c_{T}^{\pm}$ is that 

\begin{align}
|k_{*}^{\pm}| \gg k_{LIGO}
\end{align}
which is necessary for the consistency of our use of the linearly perturbed equations of motion. We can translate this into a constraint on $\gamma$ by assuming as above that $P\sim (\gamma/3)H_{0}a_{i}^{3/2}a^{-3}$ and so using equation \ref{totalcollapse} we have $k_{*}^{\pm} \sim \mp H_{0}/\gamma$ and so

\begin{align}
|\gamma| \ll {\cal O}(10^{-21})
\end{align}

\section{Perfect fluid domination}
\label{earlytimegravwavepropagation}
In this limit, we consider the evolution of perturbations on a background where the evolution is dominated by a combination of perfect fluids. It was shown in \ref{backgroundevolution} that there exist solutions where $\Lambda \sim 0$ and $P^{2} = \gamma^{2}\kappa\rho/27$ with $\gamma\ll 1$ and that these seem to be the solutions that yield a realistic cosmology. Assuming that these solutions hold then to quartic order in the small parameter $\gamma$ we have that:

\begin{widetext}
\begin{align}
\ddot{\cal H}^{\pm} &= -\omega^{2}_{\pm}(k,t){\cal H}^{\pm} -f_{\pm}(k,t)\dot{\cal H}^{\pm}
\label{fluidH} \\
\omega^{2}_{\pm}(k,t) &\equiv \bigg[  \bigg(1+ \frac{1}{9}\bigg(1+3\frac{p}{\rho}\bigg)\gamma^{2}\bigg)\bigg(\frac{k}{a}\bigg)^{2} \pm \frac{2}{3\sqrt{3}}\frac{1}{\sqrt{\kappa\rho}}\bigg(1+\frac{3p}{\rho}\bigg)\gamma^{3} \bigg(\frac{k}{a}\bigg)^{3}\bigg]  + {\cal O}(\gamma^{5}) \label{matomsq} \\
f_{\pm}(k,t) &\equiv \bigg[ \frac{(18-\gamma^{2})}{6\sqrt{3}}\sqrt{\kappa\rho} \pm \frac{(3p+\rho)\gamma(9+8\gamma^{2})}{27\rho}\bigg(\frac{k}{a}\bigg) + \frac{2}{3\sqrt{3}(\kappa\rho)^{3/2}}2\kappa(3p+\rho)\gamma^{2}\bigg(\frac{k}{a}\bigg)^{2}\nn\\
& \pm \frac{4}{9\kappa\rho^{2}}\kappa(3p+\rho)\gamma^{3}\bigg(\frac{k}{a}\bigg)^{3}\bigg] + {\cal O}(\gamma^{5})\nn\\
\end{align}
\end{widetext}

\section{Outlook}\label{outlook}

In this paper we revisited models of the Universe where the cosmological constant is allowed to vary as a result of 
a balancing torsion. Such theories potentially have fewer free parameters than General Relativity, but we need to consider
parity violating backgrounds so that they display acceptable late time phenomenology even at the level of background cosmological evolution~\cite{MZ}. 
Going beyond the homogeneous and isotropic approximation, 
the most obvious question concerns the propagating modes of the theory,
specifically gravitational waves. We found that indeed dramatic results and severe constraints arise in this respect.

We developed the required  perturbation theory within the first order formulation, taking into account that the
connection has parity-odd and -even components, with both potentially receiving zeroth order terms.
We also proposed a strategy for solving the more involved equations one has to contend with in this setting. 
We recovered the usual result for gravity waves if we assume the Einstein-Cartan theory (or solutions to our theory that reduce to it). For a theory with a pure Euler term we found a remarkable result that the linearly perturbed equations of motion leave the tensor perturbations either entirely undetermined (or fixed to vanish, if the term has a factor different from the one imposed by self-duality). This may well be hinting at the fact that gravity waves become `pure gauge' in this case (in analogy with what happens for a varying Lambda in the parity even branch of the background solutions).  

In the more general case, with a Pontryagin-type quasi-topological term, the situation is more promising. There are exotic 
effects, but these need not contradict observations in particular if we restrict ourselves to viable background solutions that may be currently indistinguishable from the standard $\Lambda CDM$ cosmological mode. 
At the level of perturbations, results will necessarily differ from General Relativity with ${\cal O}(\gamma^{2})$ deviations of the speed of tensor modes from unity.

Thus, our results are potentially very useful as a new model relating observations on the accelerating Universe (possibly implying of a non-constant deceleration parameter) and other gravitational observations. But even more originally, our conclusions may be of great value for in phenomenological quantum gravity. Modified dispersion relations are a major feature of phenomenological approaches to quantum gravity (see, for example,~\cite{GAC,MS1,MS2,Kow}).  Our work has added a layer to this approach by introducing {\it chiral} modifications to the propagation of tensor modes. It has been speculated that the concept of parity requires severe revision at the Planck scale~\cite{PlanckPar}. 
Furthermore, our results supplement existing findings regarding how parity violation in theories of gravity involving to extensions to Riemannian geometry can affect the propagation of gravitational waves (for example see \cite{Barrientos:2019awg,Conroy:2019ibo} for cases where the gravitational connection field has torsion and non-metricity respectively).

A number of open questions remain. Firstly, we have only considered tensor perturbations in this theory. It is expected that in the vector mode sector, there will be no new degrees of freedom present - as in the tensor mode case, a relic of the polynomial nature of the new $\Lambda RR$ terms in the  Lagrangian is that modifications will always only be enabled by a non-zero time derivative of $\Lambda(t)$ which concomitantly implies that time derivatives of the spin-connection perturbation will not appear. In the scalar sector, a new scalar degree of freedom $\delta\Lambda(x^{i},t)$ is expected to propagate and it will be important to see its effect on the cosmic microwave background CMB) and the growth of large scale structure.

There are also several avenues to study further observational signatures of this parity violation. The gravitational wave waveform will show deviations from General Relativity in both the amplitude and phase, due to amplitude and velocity birefringence effects, respectively, which both arise as a result of parity violation \cite{Zhao:2019xmm, Zhao:2019szi}. Some of these effects could potentially be constrained in second generation gravitational wave detectors, and it would be interesting to derive the modifications to the waveform due to these effects. In addition to observable signatures in propagation, there is potential to detect parity violation at the gravitational wave source through study of various types of astrophysical binary systems as proposed by \cite{Alexander:2017jmt, Yagi:2017zhb}. It has also been suggested that parity violation in the gravitational sector could leave distinct signatures in the CMB to be detected with future experiments (see e.g., \cite{Gluscevic_2010, Wang:2012fi,Bartolo:2017szm, Bartolo:2018elp, Inomata:2018rin, Masui:2017fzw, Shiraishi:2011st}), which could also be worth further exploration in the context of our theory. 

It is curious that the mini-superspace approximation reveals two branches with different symmetries and degrees of freedom. The non-perturbative Hamiltonian analysis of this model and various generalizations of it have been performed in \cite{Alexandrov:2021qry} and it would be interesting to explore the appearance of this branch structure within the more general results there. We have discovered that tensor mode fluctuations about certain background solutions additionally display interesting properties; in particular, the under-determination of the perturbed tensor equations of motion for a theory with a pure Euler term hints that a novel type of gauge symmetry may exist in restricted situations.

\section*{Acknowledgements}
We thank C. de Rham,  F. Hehl, T. Koivisto, M. Kopp, J. Noller, and I. Sawicki for helpful comments. TZ is funded by the European Research Council under the European Union's Seventh Framework Programme (FP7/2007-2013) / ERC Grant Agreement n. 617656 ``Theories and Models of the Dark Sector:
DM, Dark Energy and Gravity” and from
the European Structural and Investment Funds and the Czech
Ministry of Education, Youth and Sports (MSMT) (Project
CoGraDS - CZ.02.1.01/0.0/0.0/15003/0000437. JM was funded by the STFC Consolidated Grant ST/L00044X/1. The work of P.J. was supported by the funds from the European Regional Development Fund and the Czech Ministry of Education, Youth and Sports (MSMT): Project CoGraDS - CZ.02.1.01/0.0/0.0/15003/0000437.

\appendix

\section{Correct equations - no background equations were used}
For completeness we provide the position-space form of the perturbed Einstein and spin-connection equations, which take the following form:

\begin{widetext}
\begin{align}
\frac{\partial}{\partial t}\mathcal{E}^{IJ}+(H+g)\mathcal{E}^{IJ}+\big (\dot{g}+Hg-\Lambda\big ) H^{IJ}-\epsilon^{IKL}\bar{D}_{K}\mathcal{B}_{L}^{\ J}-2P\mathcal{B}^{IJ} &=0 \\
\epsilon^{JKL}\bar{D}_{K}H^{I}_{\ L}+2PH^{IJ}+\big (1-\frac{3g}{\Lambda^{2}}\dot{\Lambda}-\frac{6P}{\gamma\Lambda^{2}}\dot{\Lambda}\big )\mathcal{B}^{IJ}=\frac{3}{\Lambda^{2}}\dot{\Lambda}\big (\epsilon^{JKL}\bar{D}_{K}\mathcal{E}^{I}_{\ L}+2P\mathcal{E}^{IJ}\big )&+\frac{3}{\gamma\Lambda^{2}}\dot{\Lambda}\big (\epsilon^{JKL}\bar{D}_{K}\mathcal{B}^{I}_{\ L}-g\mathcal{E}^{IJ}\big ),\\
-\Big(\frac{\partial}{\partial t}H^{IJ}+(H-T)H^{IJ}\Big )+\big (1-\frac{3g}{\Lambda^{2}}\dot{\Lambda}-\frac{6P}{\gamma\Lambda^{2}}\dot{\Lambda}\big )\mathcal{E}^{IJ}=-\frac{3}{\Lambda^{2}}\dot{\Lambda}\big (\epsilon^{JKL}\bar{D}_{K}\mathcal{B}^{I}_{\ L}+2P\mathcal{B}^{IJ}\big )&+\frac{3}{\gamma\Lambda^{2}}\dot{\Lambda}\big (\epsilon^{JKL}\bar{D}_{K}\mathcal{E}^{I}_{\ L}+g\mathcal{B}^{IJ}\big ).
\end{align}
\end{widetext}

\section{Alternative form of the background equations of motion}
\label{backeqs}
It is possible to write down the field equations (\ref{Eeq})-(\ref{Leq}) as a system of first-order ordinary differential equations for variables $\{P,g,\Lambda,a\}$ along with a constraint equation:
\begin{widetext}
	\begin{align}
	\dot{P} &= \frac{-6 P \kappa\rho  \left(6 \gamma P g^2 +3 g \left(\Lambda+2 P^2\right)-  \Lambda\gamma P\right)+\kappa^{2}\rho  (6 P (3 g p+\gamma  P \rho )+\gamma  \Lambda\kappa (\rho -3 p))-\gamma  \kappa^{3}\rho ^2 (3 p+\rho )}{6 \kappa\rho  (6 \gamma
		 Pg+\Lambda\kappa+\rho )} \\
	\dot{g} &=-\frac{6 \left(\gamma ^2+1\right) g^2 P^2}{6 \gamma   Pg+\Lambda+\kappa\rho }-g^2+ g \gamma  P+\frac{\Lambda}{3}-\frac{\kappa}{6} (\rho+3P)\\
	\dot{\Lambda} &= \frac{2 \gamma  P \Lambda ^2 }{6 \gamma   Pg+\Lambda+\kappa\rho }\label{dotLambdaeq}\\
	\dot{a} &= a \left(\frac{6 \left(\gamma ^2+1\right) g P^2}{6 \gamma   Pg+\Lambda+\kappa\rho }+g-\gamma  P\right)\\
	\frac{\Lambda}{3} &= 
	g^{2}-P^{2} -\kappa \frac{\rho}{3}
	\end{align}
\end{widetext}
It can be checked via differentiation of the constraint equation that $\dot{\rho} = -3\frac{\dot{a}}{a}(\rho+p)$ as in the case of a perfect fluid in General Relativity. Using the constraint equation we can rewrite the evolution equation as $\dot{\Lambda} =  (2/3)\gamma P \Lambda^{2}/(g^{2}-P^{2}+2g\gamma P)$. From this perspective, the dynamics for $\Lambda$ can be seen as arising from the spin connection components $g$ and $P$.

\section{Linearly perturbed field strength and torsion}
\label{linpertor}
Central objects in the field equations (\ref{Eeq})-(\ref{Leq}) are the curvature and torsion two-forms $R^{AB}$ and $T^{A}$. Their linearly-perturbed forms around the cosmological background are found to be:

%
\begin{align}
	\delta R^{0I} &=-\epsilon^{I}_{\ph{I}JK}g{\cal B}^{J}_{\ph{J}L}\bare^{L}\bar{e}^{K}+ \bigg(\frac{1}{N}\frac{\partial}{\partial t}{\cal E}^{IJ}+\frac{1}{N}\frac{\dot{a}}{a}{\cal E}^{IJ}\bigg)\bare^{0}\bar{e}_{J}\nn\\
	&+ \bar{D}^{K}{\cal E}^{IJ}\bare_{K}\bar{e}_{J}+P{\cal E}^{I}_{\ph{I}J}\eps^{JKL}\bar{e}_{K}\bar{e}_{L}
\end{align}
\begin{align}
	\delta R^{IJ} &=    2g{\cal E}_{L}^{\ph{L}[I}\bare^{|L|}\bar{e}^{J]} 
		+ \eps^{IJ}_{\ph{IJ}K}\bigg(\bigg(\frac{1}{N}\frac{\partial}{\partial t}{\cal B}^{KL}+\frac{1}{N}\frac{\dot{a}}{a}{\cal B}^{KL}\bigg)\bare^{0}\bar{e}_{L}\nn\\
		&+(\bar{D}_{L}{\cal B}^{KM})\bare^{L}\bar{e}_{M} + P{\cal B}^{K}_{\ph{K}P}\eps^{PMN}\bare_{M}\bare_{N}\bigg)
\end{align}
\begin{align}
	\delta T^{I} &=    \left(\frac{1}{N}\frac{\partial}{\partial t} {\cal H}^{IM}+\frac{1}{N}\frac{\dot{a}}{a}{\cal H}^{IM}-{\cal E}^{IM}\right)\bare^{0}\bare_{M}\nn\\
	&+ \left(\bd^{L}{\cal H}^{IJ}+P{\cal H}^{I}_{\ph{I}K}\eps^{KLJ}+ \eps^{IJ}_{\ph{IJ}K}{\cal B}^{KL}\right)\bare_{L}\bare_{J}
\end{align}
where $\bd_{I} \equiv e^{i}_{I}\bd_{i}$ where $\bd_{i}$ is the covariant derivative according to $\bar{\omega}^{IJ}$.

\bibliographystyle{unsrtnat}
\bibliography{references}

\begin{thebibliography}{52}
\providecommand{\natexlab}[1]{#1}
\providecommand{\url}[1]{\texttt{#1}}
\expandafter\ifx\csname urlstyle\endcsname\relax
  \providecommand{\doi}[1]{doi: #1}\else
  \providecommand{\doi}{doi: \begingroup \urlstyle{rm}\Url}\fi

\bibitem[Alexander et~al.(2019{\natexlab{a}})Alexander, Cortes, Liddle,
  Magueijo, Sims, and Smolin]{alex1}
Stephon Alexander, Marina Cortes, Andrew~R. Liddle, João Magueijo, Robert
  Sims, and Lee Smolin.
\newblock {Zero-parameter extension of general relativity with a varying
  cosmological constant}.
\newblock \emph{Phys. Rev.}, D100\penalty0 (8):\penalty0 083506,
  2019{\natexlab{a}}.
\newblock \doi{10.1103/PhysRevD.100.083506}.

\bibitem[Alexander et~al.(2019{\natexlab{b}})Alexander, Cortes, Liddle,
  Magueijo, Sims, and Smolin]{alex2}
Stephon Alexander, Marina Cortes, Andrew~R. Liddle, João Magueijo, Robert
  Sims, and Lee Smolin.
\newblock {Cosmology of minimal varying Lambda theories}.
\newblock \emph{Phys. Rev.}, D100\penalty0 (8):\penalty0 083507,
  2019{\natexlab{b}}.
\newblock \doi{10.1103/PhysRevD.100.083507}.

\bibitem[Dolan(2010)]{Dolan:2009ni}
Brian~P. Dolan.
\newblock {Chiral fermions and torsion in the early Universe}.
\newblock \emph{Class. Quant. Grav.}, 27:\penalty0 095010, 2010.
\newblock \doi{10.1088/0264-9381/27/9/095010, 10.1088/0264-9381/27/24/249801}.
\newblock [Erratum: Class. Quant. Grav.27,249801(2010)].

\bibitem[Mercuri and Taveras(2009)]{Mercuri:2009zt}
Simone Mercuri and Victor Taveras.
\newblock {Interaction of the Barbero-Immirzi Field with Matter and
  Pseudo-Scalar Perturbations}.
\newblock \emph{Phys. Rev.}, D80:\penalty0 104007, 2009.
\newblock \doi{10.1103/PhysRevD.80.104007}.

\bibitem[Baekler et~al.(2011)Baekler, Hehl, and Nester]{Baekler:2010fr}
Peter Baekler, Friedrich~W. Hehl, and James~M. Nester.
\newblock {Poincare gauge theory of gravity: Friedman cosmology with even and
  odd parity modes. Analytic part}.
\newblock \emph{Phys. Rev.}, D83:\penalty0 024001, 2011.
\newblock \doi{10.1103/PhysRevD.83.024001}.

\bibitem[Poplawski(2012)]{Poplawski:2011j}
Nikodem~J. Poplawski.
\newblock {Nonsingular, big-bounce cosmology from spinor-torsion coupling}.
\newblock \emph{Phys. Rev.}, D85:\penalty0 107502, 2012.
\newblock \doi{10.1103/PhysRevD.85.107502}.

\bibitem[Schucker and Tilquin(2012)]{Schucker:2011tc}
Thomas Schucker and Andre Tilquin.
\newblock {Torsion, an alternative to the cosmological constant?}
\newblock \emph{Int. J. Mod. Phys.}, D21:\penalty0 1250089, 2012.
\newblock \doi{10.1142/S0218271812500897}.

\bibitem[Cid et~al.(2018)Cid, Izaurieta, Leon, Medina, and
  Narbona]{Cid:2017wtf}
Antonella Cid, Fernando Izaurieta, Genly Leon, Perla Medina, and Daniela
  Narbona.
\newblock {Non-minimally coupled scalar field cosmology with torsion}.
\newblock \emph{JCAP}, 1804\penalty0 (04):\penalty0 041, 2018.
\newblock \doi{10.1088/1475-7516/2018/04/041}.

\bibitem[Zlosnik et~al.(2018)Zlosnik, Urban, Marzola, and
  Koivisto]{Zlosnik:2018qvg}
Tom Zlosnik, Federico Urban, Luca Marzola, and Tomi Koivisto.
\newblock {Spacetime and dark matter from spontaneous breaking of Lorentz
  symmetry}.
\newblock \emph{Class. Quant. Grav.}, 35\penalty0 (23):\penalty0 235003, 2018.
\newblock \doi{10.1088/1361-6382/aaea96}.

\bibitem[Flanagan(2004)]{Flanagan:2003rb}
Eanna~E. Flanagan.
\newblock {Palatini form of 1/R gravity}.
\newblock \emph{Phys. Rev. Lett.}, 92:\penalty0 071101, 2004.
\newblock \doi{10.1103/PhysRevLett.92.071101}.

\bibitem[Sotiriou(2006)]{Sotiriou:2005hu}
Thomas~P. Sotiriou.
\newblock {Unification of inflation and cosmic acceleration in the Palatini
  formalism}.
\newblock \emph{Phys. Rev.}, D73:\penalty0 063515, 2006.
\newblock \doi{10.1103/PhysRevD.73.063515}.

\bibitem[Bauer and Demir(2011)]{Bauer:2010jg}
Florian Bauer and Durmus~A. Demir.
\newblock {Higgs-Palatini Inflation and Unitarity}.
\newblock \emph{Phys. Lett.}, B698:\penalty0 425--429, 2011.
\newblock \doi{10.1016/j.physletb.2011.03.042}.

\bibitem[Farnsworth et~al.(2017)Farnsworth, Lehners, and
  Qiu]{Farnsworth:2017wzr}
Shane Farnsworth, Jean-Luc Lehners, and Taotao Qiu.
\newblock {Spinor driven cosmic bounces and their cosmological perturbations}.
\newblock \emph{Phys. Rev.}, D96\penalty0 (8):\penalty0 083530, 2017.
\newblock \doi{10.1103/PhysRevD.96.083530}.

\bibitem[Addazi et~al.(2019)Addazi, Chen, and Marciano]{Addazi:2017qus}
Andrea Addazi, Pisin Chen, and Antonino Marciano.
\newblock {Emergent inflation from a Nambu--Jona-Lasinio mechanism in gravity
  with non-dynamical torsion}.
\newblock \emph{Eur. Phys. J.}, C79\penalty0 (4):\penalty0 297, 2019.
\newblock \doi{10.1140/epjc/s10052-019-6803-7}.

\bibitem[Magueijo and Z{\l}o{\'s}nik(2019)]{MZ}
Jo{\~a}o Magueijo and Tom Z{\l}o{\'s}nik.
\newblock {Parity violating Friedmann Universes}.
\newblock \emph{Phys. Rev.}, D100\penalty0 (8):\penalty0 084036, 2019.
\newblock \doi{10.1103/PhysRevD.100.084036}.

\bibitem[Lazar and Hehl(2010)]{spiral}
Markus Lazar and Friedrich~W. Hehl.
\newblock {Cartan's spiral staircase in physics and, in particular, in the
  gauge theory of dislocations}.
\newblock \emph{Found. Phys.}, 40:\penalty0 1298--1325, 2010.
\newblock \doi{10.1007/s10701-010-9440-4}.

\bibitem[Henneaux and Teitelboim(1989)]{Henneaux:1989zc}
M.~Henneaux and C.~Teitelboim.
\newblock {The Cosmological Constant and General Covariance}.
\newblock \emph{Phys. Lett.}, B222:\penalty0 195--199, 1989.
\newblock \doi{10.1016/0370-2693(89)91251-3}.

\bibitem[Jirousek and Vikman(2019)]{Jirousek:2018ago}
Pavel Jirousek and Alexander Vikman.
\newblock {New Weyl-invariant vector-tensor theory for the cosmological
  constant}.
\newblock \emph{JCAP}, 1904:\penalty0 004, 2019.
\newblock \doi{10.1088/1475-7516/2019/04/004}.

\bibitem[Hammer et~al.(2020)Hammer, Jirousek, and Vikman]{Hammer:2020dqp}
Katrin Hammer, Pavel Jirousek, and Alexander Vikman.
\newblock {Axionic cosmological constant}.
\newblock 2020.

\bibitem[Alexander et~al.(2019{\natexlab{c}})Alexander, Magueijo, and
  Smolin]{Alexander:2018djy}
Stephon Alexander, Joao Magueijo, and Lee Smolin.
\newblock {The Quantum Cosmological Constant}.
\newblock \emph{Symmetry}, 11\penalty0 (9):\penalty0 1130, 2019{\natexlab{c}}.
\newblock \doi{10.3390/sym11091130}.

\bibitem[Magueijo and Smolin(2018)]{LeeJ}
Jo\~{a}o Magueijo and Lee Smolin.
\newblock {A Universe that does not know the time}.
\newblock 2018.
\newblock \doi{10.3390/universe5030084}.

\bibitem[Smolin and Soo(1995)]{chopin}
Lee Smolin and Chopin Soo.
\newblock {The Chern-Simons invariant as the natural time variable for
  classical and quantum cosmology}.
\newblock \emph{Nucl. Phys.}, B449:\penalty0 289--316, 1995.
\newblock \doi{10.1016/0550-3213(95)00222-E}.

\bibitem[Gott and Li(1998)]{Gott:1997pm}
J.~Richard Gott, III and Li-Xin Li.
\newblock {Can the universe create itself?}
\newblock \emph{Phys. Rev.}, D58:\penalty0 023501, 1998.
\newblock \doi{10.1103/PhysRevD.58.023501}.

\bibitem[Afshordi et~al.(2007)Afshordi, Chung, and Geshnizjani]{cuscaton}
Niayesh Afshordi, Daniel J.~H. Chung, and Ghazal Geshnizjani.
\newblock {Cuscuton: A Causal Field Theory with an Infinite Speed of Sound}.
\newblock \emph{Phys. Rev.}, D75:\penalty0 083513, 2007.
\newblock \doi{10.1103/PhysRevD.75.083513}.

\bibitem[Aros et~al.(2000)Aros, Contreras, Olea, Troncoso, and
  Zanelli]{Aros:1999id}
Rodrigo Aros, Mauricio Contreras, Rodrigo Olea, Ricardo Troncoso, and Jorge
  Zanelli.
\newblock {Conserved charges for gravity with locally AdS asymptotics}.
\newblock \emph{Phys. Rev. Lett.}, 84:\penalty0 1647--1650, 2000.
\newblock \doi{10.1103/PhysRevLett.84.1647}.

\bibitem[Miskovic and Olea(2009)]{Miskovic:2009bm}
Olivera Miskovic and Rodrigo Olea.
\newblock {Topological regularization and self-duality in four-dimensional
  anti-de Sitter gravity}.
\newblock \emph{Phys. Rev.}, D79:\penalty0 124020, 2009.
\newblock \doi{10.1103/PhysRevD.79.124020}.

\bibitem[Ma and Bertschinger(1995)]{Ma:1995ey}
Chung-Pei Ma and Edmund Bertschinger.
\newblock {Cosmological perturbation theory in the synchronous and conformal
  Newtonian gauges}.
\newblock \emph{Astrophys. J.}, 455:\penalty0 7--25, 1995.
\newblock \doi{10.1086/176550}.

\bibitem[{Planck Collaboration} et~al.(2015){Planck Collaboration}, {Ade},
  {Aghanim}, {Arnaud}, {Ashdown}, {Aumont}, {Baccigalupi}, {Banday},
  {Barreiro}, {Bartlett}, and et~al.]{PlanckCollaboration2015}
{Planck Collaboration}, P.~A.~R. {Ade}, N.~{Aghanim}, M.~{Arnaud},
  M.~{Ashdown}, J.~{Aumont}, C.~{Baccigalupi}, A.~J. {Banday}, R.~B.
  {Barreiro}, J.~G. {Bartlett}, and et~al.
\newblock {Planck 2015 results. XIII. Cosmological parameters}.
\newblock \emph{ArXiv e-prints}, February 2015.

\bibitem[Carroll and Lim(2004)]{Carroll:2004ai}
Sean~M. Carroll and Eugene~A. Lim.
\newblock {Lorentz-violating vector fields slow the universe down}.
\newblock \emph{Phys. Rev. D}, 70:\penalty0 123525, 2004.
\newblock \doi{10.1103/PhysRevD.70.123525}.

\bibitem[Barrientos et~al.(2019)Barrientos, Cordonier-Tello, Corral, Izaurieta,
  Medina, Rodr{\'\i}guez, and Valdivia]{Barrientos:2019awg}
Jos{\'e} Barrientos, Fabrizio Cordonier-Tello, Crist{\'o}bal Corral, Fernando
  Izaurieta, Perla Medina, Eduardo Rodr{\'\i}guez, and Omar Valdivia.
\newblock {Luminal Propagation of Gravitational Waves in Scalar-tensor
  Theories: The Case for Torsion}.
\newblock \emph{Phys. Rev.}, D100\penalty0 (12):\penalty0 124039, 2019.
\newblock \doi{10.1103/PhysRevD.100.124039}.

\bibitem[Baker et~al.(2017)Baker, Bellini, Ferreira, Lagos, Noller, and
  Sawicki]{Johan}
T.~Baker, E.~Bellini, P.~G. Ferreira, M.~Lagos, J.~Noller, and I.~Sawicki.
\newblock {Strong constraints on cosmological gravity from GW170817 and GRB
  170817A}.
\newblock \emph{Phys. Rev. Lett.}, 119\penalty0 (25):\penalty0 251301, 2017.
\newblock \doi{10.1103/PhysRevLett.119.251301}.

\bibitem[Creminelli and Vernizzi(2017)]{Creminelli_2017}
Paolo Creminelli and Filippo Vernizzi.
\newblock Dark energy after gw170817 and grb170817a.
\newblock \emph{Physical Review Letters}, 119\penalty0 (25), Dec 2017.
\newblock ISSN 1079-7114.
\newblock \doi{10.1103/physrevlett.119.251302}.
\newblock URL \url{http://dx.doi.org/10.1103/PhysRevLett.119.251302}.

\bibitem[Sakstein and Jain(2017)]{Sakstein_2017}
Jeremy Sakstein and Bhuvnesh Jain.
\newblock Implications of the neutron star merger gw170817 for cosmological
  scalar-tensor theories.
\newblock \emph{Physical Review Letters}, 119\penalty0 (25), Dec 2017.
\newblock ISSN 1079-7114.
\newblock \doi{10.1103/physrevlett.119.251303}.
\newblock URL \url{http://dx.doi.org/10.1103/PhysRevLett.119.251303}.

\bibitem[Ezquiaga and Zumalac{\'a}rregui(2017)]{Ezquiaga_2017}
Jose~Mar{\'\i}a Ezquiaga and Miguel Zumalac{\'a}rregui.
\newblock Dark energy after gw170817: Dead ends and the road ahead.
\newblock \emph{Physical Review Letters}, 119\penalty0 (25), Dec 2017.
\newblock ISSN 1079-7114.
\newblock \doi{10.1103/physrevlett.119.251304}.
\newblock URL \url{http://dx.doi.org/10.1103/PhysRevLett.119.251304}.

\bibitem[Amelino-Camelia(2002)]{GAC}
Giovanni Amelino-Camelia.
\newblock {Relativity in space-times with short distance structure governed by
  an observer independent (Planckian) length scale}.
\newblock \emph{Int. J. Mod. Phys.}, D11:\penalty0 35--60, 2002.
\newblock \doi{10.1142/S0218271802001330}.

\bibitem[Magueijo and Smolin(2002)]{MS1}
Joao Magueijo and Lee Smolin.
\newblock {Lorentz invariance with an invariant energy scale}.
\newblock \emph{Phys. Rev. Lett.}, 88:\penalty0 190403, 2002.
\newblock \doi{10.1103/PhysRevLett.88.190403}.

\bibitem[Magueijo and Smolin(2003)]{MS2}
Joao Magueijo and Lee Smolin.
\newblock {Generalized Lorentz invariance with an invariant energy scale}.
\newblock \emph{Phys. Rev.}, D67:\penalty0 044017, 2003.
\newblock \doi{10.1103/PhysRevD.67.044017}.

\bibitem[Kowalski-Glikman and Nowak(2002)]{Kow}
J.~Kowalski-Glikman and S.~Nowak.
\newblock {Doubly special relativity theories as different bases of kappa
  Poincare algebra}.
\newblock \emph{Phys. Lett.}, B539:\penalty0 126--132, 2002.
\newblock \doi{10.1016/S0370-2693(02)02063-4}.

\bibitem[Arzano et~al.(2018)Arzano, Gubitosi, and Magueijo]{PlanckPar}
Michele Arzano, Giulia Gubitosi, and Joao Magueijo.
\newblock {Parity at the Planck scale}.
\newblock \emph{Phys. Lett.}, B781:\penalty0 510--516, 2018.
\newblock \doi{10.1016/j.physletb.2018.04.027}.

\bibitem[Conroy and Koivisto(2019)]{Conroy:2019ibo}
Aindri{\'u} Conroy and Tomi Koivisto.
\newblock {Parity-Violating Gravity and GW170817 in Non-Riemannian Cosmology}.
\newblock 2019.
\newblock \doi{10.1088/1475-7516/2019/12/016}.

\bibitem[Zhao et~al.(2019{\natexlab{a}})Zhao, Zhu, Qiao, and
  Wang]{Zhao:2019xmm}
Wen Zhao, Tao Zhu, Jin Qiao, and Anzhong Wang.
\newblock {Waveform of gravitational waves in the general parity-violating
  gravities}.
\newblock 2019{\natexlab{a}}.

\bibitem[Zhao et~al.(2019{\natexlab{b}})Zhao, Liu, Wen, Zhu, Wang, Hu, and
  Zhou]{Zhao:2019szi}
Wen Zhao, Tan Liu, Linqing Wen, Tao Zhu, Anzhong Wang, Qian Hu, and Cong Zhou.
\newblock {Model-independent test of the parity symmetry of gravity with
  gravitational waves}.
\newblock 2019{\natexlab{b}}.

\bibitem[Alexander and Yunes(2018)]{Alexander:2017jmt}
Stephon~H. Alexander and Nicolas Yunes.
\newblock {Gravitational wave probes of parity violation in compact binary
  coalescences}.
\newblock \emph{Phys. Rev.}, D97\penalty0 (6):\penalty0 064033, 2018.
\newblock \doi{10.1103/PhysRevD.97.064033}.

\bibitem[Yagi and Yang(2018)]{Yagi:2017zhb}
Kent Yagi and Huan Yang.
\newblock {Probing Gravitational Parity Violation with Gravitational Waves from
  Stellar-mass Black Hole Binaries}.
\newblock \emph{Phys. Rev.}, D97\penalty0 (10):\penalty0 104018, 2018.
\newblock \doi{10.1103/PhysRevD.97.104018}.

\bibitem[Gluscevic and Kamionkowski(2010)]{Gluscevic_2010}
Vera Gluscevic and Marc Kamionkowski.
\newblock Testing parity-violating mechanisms with cosmic microwave background
  experiments.
\newblock \emph{Physical Review D}, 81\penalty0 (12), Jun 2010.
\newblock ISSN 1550-2368.
\newblock \doi{10.1103/physrevd.81.123529}.
\newblock URL \url{http://dx.doi.org/10.1103/PhysRevD.81.123529}.

\bibitem[Wang et~al.(2013)Wang, Wu, Zhao, and Zhu]{Wang:2012fi}
Anzhong Wang, Qiang Wu, Wen Zhao, and Tao Zhu.
\newblock {Polarizing primordial gravitational waves by parity violation}.
\newblock \emph{Phys. Rev.}, D87\penalty0 (10):\penalty0 103512, 2013.
\newblock \doi{10.1103/PhysRevD.87.103512}.

\bibitem[Bartolo and Orlando(2017)]{Bartolo:2017szm}
Nicola Bartolo and Giorgio Orlando.
\newblock {Parity breaking signatures from a Chern-Simons coupling during
  inflation: the case of non-Gaussian gravitational waves}.
\newblock \emph{JCAP}, 1707:\penalty0 034, 2017.
\newblock \doi{10.1088/1475-7516/2017/07/034}.

\bibitem[Bartolo et~al.(2019)Bartolo, Orlando, and Shiraishi]{Bartolo:2018elp}
Nicola Bartolo, Giorgio Orlando, and Maresuke Shiraishi.
\newblock {Measuring chiral gravitational waves in Chern-Simons gravity with
  CMB bispectra}.
\newblock \emph{JCAP}, 1901\penalty0 (01):\penalty0 050, 2019.
\newblock \doi{10.1088/1475-7516/2019/01/050}.

\bibitem[Inomata and Kamionkowski(2019)]{Inomata:2018rin}
Keisuke Inomata and Marc Kamionkowski.
\newblock {Chiral photons from chiral gravitational waves}.
\newblock \emph{Phys. Rev. Lett.}, 123\penalty0 (3):\penalty0 031305, 2019.
\newblock \doi{10.1103/PhysRevLett.123.031305}.

\bibitem[Masui et~al.(2017)Masui, Pen, and Turok]{Masui:2017fzw}
Kiyoshi~Wesley Masui, Ue-Li Pen, and Neil Turok.
\newblock {Two- and Three-Dimensional Probes of Parity in Primordial Gravity
  Waves}.
\newblock \emph{Phys. Rev. Lett.}, 118\penalty0 (22):\penalty0 221301, 2017.
\newblock \doi{10.1103/PhysRevLett.118.221301}.

\bibitem[Shiraishi et~al.(2011)Shiraishi, Nitta, and
  Yokoyama]{Shiraishi:2011st}
Maresuke Shiraishi, Daisuke Nitta, and Shuichiro Yokoyama.
\newblock {Parity Violation of Gravitons in the CMB Bispectrum}.
\newblock \emph{Prog. Theor. Phys.}, 126:\penalty0 937--959, 2011.
\newblock \doi{10.1143/PTP.126.937}.

\bibitem[Alexandrov et~al.(2021)Alexandrov, Speziale, and
  Zlosnik]{Alexandrov:2021qry}
Sergei Alexandrov, Simone Speziale, and Tom Zlosnik.
\newblock {Canonical structure of minimal varying \ensuremath{\Lambda}
  theories}.
\newblock \emph{Class. Quant. Grav.}, 38\penalty0 (17):\penalty0 175011, 2021.
\newblock \doi{10.1088/1361-6382/ac1852}.

\end{thebibliography}

\end{document}